\documentclass[traditabstract]{aa}
\usepackage{graphicx}
\usepackage{txfonts}
\usepackage{natbib}
\usepackage{enumitem}
\usepackage{rotating}
\usepackage{pdflscape}

\bibpunct{(}{)}{;}{a}{}{,}

 \newcommand{\mic}{$\mu$m}
 \newcommand{\md}{$M_{dust}$}
 \newcommand{\ms}{$M_{star}$}

 \newcommand{\msun}{$M_\odot$}

\begin{document}

\title{Characterizing elusive, faint dusty star-forming galaxies: a lensed,
optically undetected ALMA galaxy at $z$$\sim$3.3} 

\author{
P.~Santini\inst{1}
\and
 M.~Castellano\inst{1} 
 \and
 A.~Fontana\inst{1}   
 \and
 E.~Merlin\inst{1}  
 \and
 R.~Maiolino\inst{2,3} 
 \and
 C.~Mason\inst{4,5}   
 \and
 A.~Mignano\inst{6}  
 \and
 S.~Pilo\inst{1}  
 \and
 R.~Amorin\inst{1,2,3}
 \and
  S.~Berta\inst{7}  
 \and
 N.~Bourne\inst{8}  
 \and
 F.~Calura\inst{9}  
 \and
 E.~Daddi\inst{10}
 \and
 D.~Elbaz\inst{10}  
 \and
 A.~Grazian\inst{1}  
  \and
  M.~Magliocchetti\inst{11}   
 \and
 M.~J.~Micha{\l}owski\inst{8}    
 \and
 L.~Pentericci\inst{1}
 \and
 F.~Pozzi\inst{12,9}   
 \and
 G.~Rodighiero\inst{13}   
 \and
 C.~Schreiber\inst{14}  
 \and
 R.~Valiante\inst{1}  
}

  \offprints{P. Santini, \email{paola.santini@oa-roma.inaf.it}}

\institute{INAF - Osservatorio Astronomico di Roma, via di Frascati 33, 00078 Monte Porzio Catone, Italy
 \and Cavendish Laboratory, University of Cambridge, 19 J. J. Thomson
 Ave., Cambridge CB3 0HE, UK
 \and Kavli Institute for Cosmology, University of Cambridge, Madingley Road, Cambridge CB3 0HA, UK
 \and Department of Physics, University of California, Santa Barbara,
 CA, 93106-9530, USA 
 \and Department of Physics and Astronomy, UCLA, Los Angeles, CA,
 90095-1547, USA
\and INAF - Istituto di Radioastronomia, via P. Gobetti 101, I-40129
Bologna, Italy
\and ste\_atreb@yahoo.it
 \and Institute for Astronomy, University of Edinburgh, Royal Observatory,
 Edinburgh, EH9 3HJ, U.K
 \and INAF - Osservatorio Astronomico di Bologna, via Ranzani 1, 40127 Bologna, Italy
 \and Laboratoire AIM, CEA/DSM-CNRS-Universit{\'e} Paris Diderot, IRFU/Service d'Astrophysique, B\^at.709, CEA-Saclay, 91191 Gif-sur-Yvette Cedex, France
 \and INAF – IAPS, via Fosso del Cavaliere 100, 00133 Roma, Italy
 \and Dipartimento di Astronomia, Universit{\`a} di Bologna, via
 Ranzani 1, 40127 Bologna, Italy
 \and Dipartimento di Astronomia, Universit\`à di Padova, Vicolo
 dell’Osservatorio 3, 35122 Padova, Italy
\and Leiden Observatory, Leiden University, NL-2300 RA Leiden, The Netherlands
}

   \date{Received .... ; accepted ....}
   \titlerunning{Characterizing elusive, faint dusty star-forming
     galaxies}

   \abstract{ We present the serendipitous ALMA detection of a faint
     submillimeter galaxy (SMG) lensed by a foreground
     $z$$\sim$1
     galaxy.  By optimizing the source detection to deblend the
     system, we accurately build the full spectral energy distribution
     of the distant galaxy from the I814 band to radio wavelengths. It
     is extremely red, with a I--K colour larger than 2.5. We estimate
     a photometric redshift of 3.28 and determine the physical
     parameters. The distant galaxy turns out to be magnified by the
     foreground lens by a factor of $\sim$1.5,
     which implies an intrinsic K$_s$-band
     magnitude of $\sim$24.5,
     a submillimeter flux at 870\mic~of $\sim$2.5
     mJy and a SFR of $\sim
     150-300M_\odot/yr$, depending on the adopted tracer.  These
     values place our source towards the faint end of the distribution
     of observed SMGs, and in particular among the still few faint
     SMGs with a fully characterized spectral energy distribution,
     which allows us not only to accurately estimate its redshift, but
     also to measure its stellar mass and other physical properties.
     The galaxy studied in this work is a representative of the
     population of faint SMGs, of which only few objects are known to
     date, that are undetected in optical and therefore are not
     typically accounted for when measuring the cosmic star formation
     history (SFH).  This faint galaxy population thus likely
     represents an important and missing piece in our understanding of
     the cosmic SFH. Its observation and characterization is of major
     importance to achieve a solid picture of galaxy evolution.  }


\maketitle

\section{Introduction}\label{sec:intro}

The cosmic star formation history (SFH) is a key observable to
understand galaxy evolution and constrain theoretical models. Since
the seminal works of \cite{madau96} and \cite{lilly96}, much
effort has been made to constrain its shape (see the review of
\citealt{madau14}).  The launch of {\it Herschel} and the advent of ALMA
have allowed us to measure the cosmic SFH out to $z\sim 3$ free of
uncertain dust extinction corrections (e.g. \citealt{burgarella13,
  dunlop16,bouwens16}, but see also \citealt{bourne16} for a recent
result based on SCUBA-2).  However, because of limited ALMA
observations (combined with its small field of view) and confusion in
the far-infrared (FIR) images, at high redshift we still mostly rely
on the rest-frame UV observations, which must be corrected for dust
absorption. Nevertheless, at the peak of the cosmic star formation
(SF) activity, the power emitted by young stars in the IR (through
dust reprocessing) is an order of magnitude higher than that emitted
in the UV \citep{madau14,dunlop16}.  Incorrect dust corrections could
result in an incorrect overall picture of the SFH of the Universe
\citep[e.g.][]{castellano14}.

Another serious issue affecting the measure of the cosmic SFH is
whether our census is complete. In other words, are we counting all
star-forming galaxies? Or are we missing a fraction of them?
Observational evidence seems to point towards the latter scenario (see
e.g. the two extremely red galaxies found by \citealt{caputi14} or
the HIEROs galaxies presented by \citealt{wang16}).  Most of our
knowledge is based on galaxies selected in the UV, optical, or near-IR
(submm and mm samples are also available, but they are highly
incomplete, see \citealt{casey14} for a collection of results from the
literature). However, according to \cite{viero13}, K-detected galaxies
account for only $\sim$70\% of the submillimeter background (but see
also \citealt{viero15}, who claimed that most of it, if not all, can be
accounted for by low-mass sources).  Since $\sim$95\% of the
submillimeter background is contributed by dusty star-forming
galaxies, it is very likely that we are missing a substantial
contribution to the cosmic SFH. As candidates for the missing
fraction, \cite{viero13} indicated low-mass faint sources and
dust-obscured galaxies.  Identifying galaxies undetected in the
optical/near-IR bands is also extremely important to address the still
open issue regarding the missing mass (the observed stellar mass is
lower than what is obtained by integrating the SFH at $z\lesssim 2-3$;
e.g. \citealt{santini12a}, \citealt{madau14}, \citealt{grazian15}).

A population of IR bright galaxies that are often undetected at
optical/near-IR wavelengths are the so-called submillimeter galaxies
\citep[SMGs, e.g.][]{hughes98,chapman05,greve05,pope06,tacconi08}, a
heterogeneous \citep{magnelli12a,dacunha15} population of  massive
($10^{10}-10^{11}M_\odot$), strongly star-forming,
heavily dust-obscured galaxies historically selected from their submm
flux.  They  contribute substantially to the cosmic SFH at
$z\sim 2 - 3$ \citep{chapman05,michalowski10a} and are responsible for almost the
entire dust-obscured SF \citep{barger12,casey13,casey14}.

The brightest SMGs, making hundreds or thousands of solar masses per
year, are quite rare and their overall contribution to the cosmic SFH
is modest \citep[e.g.][]{leborgne09,bethermin11,swinbank14}.
Typically, observed SMGs have submm fluxes higher than a few mJy
(e.g. $S_{860\mu m}$$>$3
mJy for the SMA survey of \citealt{barger12} in GOODS-N, or $S_{870\mu
  m}$$>$4.4 mJy for the LESS survey with the LABOCA camera in the
ECDF, \citealt{weiss09}), although a significant fraction of them has
been resolved into multiple sources by ALMA observations \citep[ALESS
survey,][]{hodge13}. However, the SMG population is dominated by
fainter sources (see \citealt{casey14} and references therein,
\citealt{dunlop16}, \citealt{aravena16b}).  According to \cite{ono14},
faint SMGs, defined as having 1.2~mm fluxes between 0.1 and 1~mJy and
SFRs between 30 and 300 $M_\odot/yr$, contribute nearly half of the
submm extragalactic background (with the other half being accounted
for by even fainter sources).  Moreover, their contribution to the SFH
is at least as much as that of bright ($S_{870\mu m}$$>$4
mJy) SMGs \citep{yamaguchi16}.  However, despite their abundance and
significant levels of SF, many faint SMGs may not be included in the
cosmic SFH, as they are undetected at shorter wavelengths
\citep{hsu16}.  \cite{yamaguchi16} estimated that the contribution of
these optically undetected galaxies to the infrared SFH at
$0.9<z<3.6$
can be as large as 10\%.  A thorough investigation of this elusive
population is therefore of major importance for completing the census
of star-forming galaxies and reaching a full understanding of galaxy
evolution.

Faint SMGs are being detected thanks to ALMA capabilities
\citep{ono14,yamaguchi16,dunlop16,fujimoto16,oteo16} and to
amplification of their flux through gravitational lensing effects
\citep{chen14,hsu16}.  Gravitational lensing has recently been revealed to
be a very powerful tool to detect sources below the sensitivity limit
of current instrumentation
\citep[e.g.][]{treu10b,vanderwel13,amorin14,watson15,vanzella16,castellano16c}.

Although faint SMGs are being observed, a full (optical-to-submm)
spectral energy distribution (SED) characterization of these sources
is still missing, with few exceptions
(\citealt{simpson14,dacunha15,dunlop16,bouwens16,yamaguchi16,koprowski16},
especially in the deepest fields like the HUDF) .  We present here a
serendipitous ALMA detection of a lensed, optically undetected
$z$$\sim$3.3
faint SMG (demagnified flux $S_{870\mu
  m}$$\sim$2.5 mJy, intrinsic SFR of $\sim 150-300M_\odot/yr$),
amplified by a $z$$\sim$1
galaxy, with full characterization of its SED.  Galaxies like this one
are typically not included in the census of star-forming galaxies.
This work is an attempt to uncover the physical properties of this
elusive, but important, class of sources by adding one representative
to the still sparse population of known faint, dusty, star-forming
galaxies.

The paper is organized as follows. In Sects. \ref{sec:targetsel} -
\ref{sec:phot} we describe the target identification, the ALMA
observations and how we measure the photometry in the optical and
near-infrared (NIR) bands. In Sect.~\ref{sec:results} we present the
physical properties of the studied source. Finally, in
Sect.~\ref{sec:disc} we discuss and summarize our results.  In the
following, we adopt the $\Lambda$-CDM
concordance cosmological model (H$_0$
= 70 km/s/Mpc, $\Omega_M
= 0.3$ and $\Omega_{\Lambda} =
0.7$) and a \cite{chabrier03} IMF.  All magnitudes are in the AB
system.

\section{Target identification} \label{sec:targetsel}

The faint SMG studied in the present work was detected serendipitously
by analysing a sample of sources in the COSMOS field in detail 
\citep{scoville07} that shows a very high dust mass compared to their
stellar mass.

Stellar masses were computed by performing a standard SED fitting with
\cite{bc03} templates assuming exponentially declining SF histories
(see details in Table 1 of \citealt{santini15}, method 6a$_\tau$) on
the photometry from the UltraVISTA-DR1 $\rm K_s$--selected catalogue of
\cite{muzzin13} (30 photometric bands). Dust masses were inferred by
using the \cite{draineli07} model (with the same technique as
adopted by \citealt{santini14}) to fit the MIPS 24\mic~ \citep{lefloch09} and
{\it Herschel} photometry from the PEP \citep{lutz11} and HerMES
\citep{oliver12} surveys with PACS and SPIRE instruments, respectively
(see \citealt{lutz11}, and \citealt{berta11} for a description of PACS
catalogues and \citealt{roseboom10}, and \citealt{roseboom12} for SPIRE
catalogues, both based on prior knowledge of 24\mic~positions).

In particular, we focused our attention on a galaxy at spectroscopic
redshift of $z$=1.1458 (from the Nov. 2015 zCOSMOS release,
\citealt{lilly09}), located at RA=10:01:38.48 DEC=+02:37:35.03, with
log\ms/\msun$=$10.3 and an apparent log\md/\msun$=$9.7, which happened
to be observed by ALMA. While such high dust-to-stellar mass ratios
have been observed by a previous study \citep{pappalardo16}, ALMA
observations demonstrate that at least part of the FIR flux is
associated with a different source, 2 arcsec apart from the
optically detected galaxy, located at RA=10:01:38.547
DEC=+02:37:36.70, which was undetected in the \cite{muzzin13} catalogue and even
in the recent catalogue by \cite{laigle16} that is based on the UltraVISTA-DR2
data release.

In the following, we refer to these two galaxies as the 
optically detected and the ALMA-detected source, respectively. The latter
will also be referred to as SMG, using the purely observational
definition of the term.

\section{ALMA observations and data  analysis}\label{sec:alma}

ALMA archival observations of this sky region were carried out as part
of the Cycle 2 project 2013.1.00034.S (PI N. Scoville), split into two
runs (2014 July and December). The 12 m antennae (32 and 39) were distributed in compact configuration, with
baselines ranging from $\sim$150 m to 1.09 km.  The spectral
configuration covered a 8 GHz band centred around 343.463 GHz (Band
7). Each of the four 2 GHz spectral windows was sampled in 128
channels.  The on-source integration time was 242 seconds.  As bandpass
calibrators, J1058+0133 and J0825+0309 were observed for the runs in
July and December, respectively. The flux calibrators were Titan and
J1037-295, respectively.  J1010-0200 and J1008+0621 were used as phase
calibrators.  The phases were centred at the position of the
optically detected source described in the previous section.

\begin{figure}[!t]
\resizebox{\hsize}{!}{\includegraphics[angle=0]{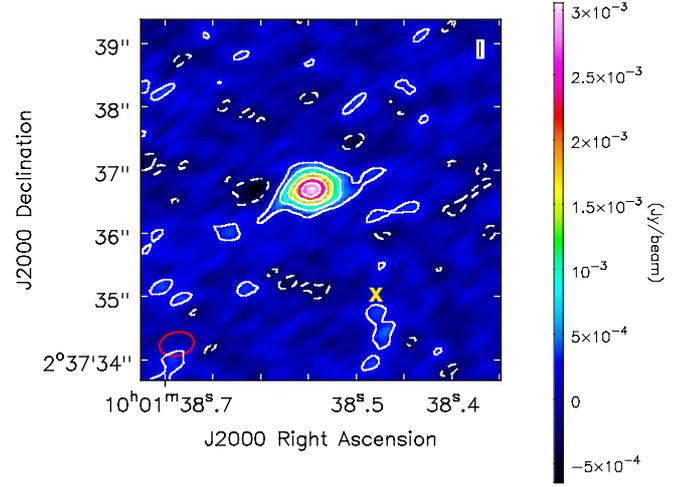}}
\caption{Continuum in ALMA Band 7 averaged over the four spectral
  windows. Solid lines show the 2, 5, 10, 15, and 20$\sigma$ contours,
  while dashed lines show negative fluctuations at $-2\sigma$. The
  beam is shown by the red ellipse in the bottom left corner. The
  yellow 'X' shows the position of the optically detected galaxy.}
\label{fig:almacont}
\end{figure}

Observations were calibrated using the CASA pipeline version number
31667 (Pipeline-Cycle2-R1-B).  The field was imaged using the {\it
  clean} task of CASA down to 0.3 mJy ($\sim$2-3$\times$ the thermal
noise) with mode {\it 'mfs'} \citep{rau11} and weight {\it 'briggs'},
using CASA version 4.5.1.  The RMS is 0.14~mJy/beam and the beam has
a size of 0.54$\times$0.37~arcsec and a position angle of -77.92 deg. The
continuum image is shown in Fig.~\ref{fig:almacont}.

\begin{figure*}[!t]
\resizebox{\hsize}{!}{\includegraphics[angle=0]{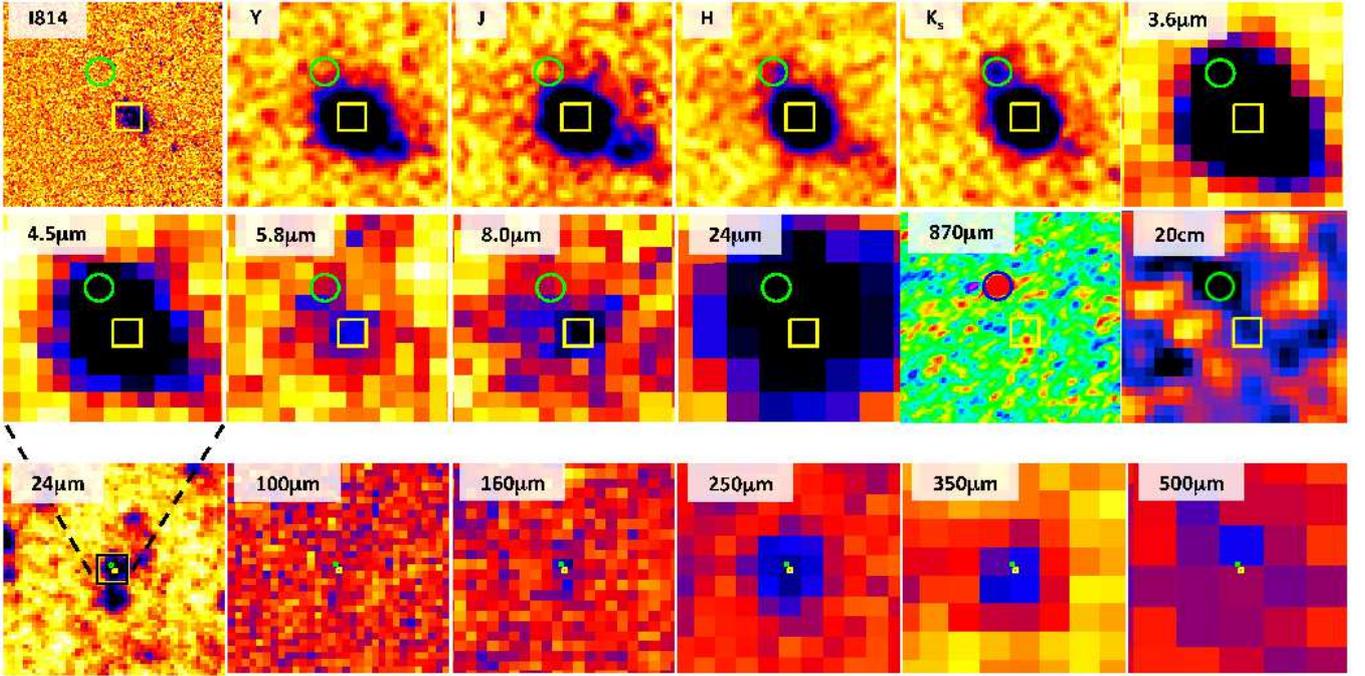}}
\caption{Image thumbnails in a region of 8x8 arcsec showing the two
  galaxies. Images are, from left to right and from top to
  bottom, HST/I band, UltraVISTA-DR3 Y, J, H and K$_s$ bands, IRAC
  CH1, CH2, CH3 and CH4, MIPS 24\mic, ALMA Band 7 at $\sim 870$\mic~
  and VLA 20cm, as indicated by the labels. In the bottom row we show
  the MIPS 24\mic, PACS 100 and 160\mic, SPIRE 250, 350 and
  500\mic~stamps in a more extended region of 60x60 arcsec.  The green
  (blue in the ALMA stamp) circle indicates the position of the
  ALMA-detected SMG and the yellow box shows that of the
  optically detected source.  The cuts in the H and K$_s$ UltraVISTA
  stamps have been optimized to show the distant galaxy. }
\label{fig:stamps}
\end{figure*}

A very bright source, well above the 20$\sigma$ level, is clearly
visible 2 arcsec apart from the position of the optically detected
galaxy, at RA=10:01:38.547 DEC=+02.37.36.70. We note that the distance
between the two sources cannot be explained by ALMA astrometric
accuracy\footnote{https://help.almascience.org/index.php?/Knowledgebase/Article/\\View/319/6/what-is-the-astrometric-accuracy-of-alma},
which, for our source, is of the order of a few $10^{-2}$ arcsec
\citep[see also][]{dunlop16}. As a further confirmation, we checked
that the astrometric uncertainty on the position derived from the maps
of the phase calibrators is lower than $10^{-2}$ arcsec.  Finally, the
ALMA position is consistent with the VLA position of a
radio-identified source (see Fig.~\ref{fig:stamps} and
Sect.~\ref{sec:results}).

To avoid being affected by issues associated with the cleaning
process, we measured the flux directly on the visibilities using the
{\it uvmodelfit} task of CASA, assuming a Gaussian model. The
integrated flux over the source is $3.90 \pm 0.41$ mJy (we added 10\%
of the flux in quadrature to the RMS to take the error in
the calibration into account).  The semimajor axes are $0.24 \pm 0.04$ and
$0.21 \pm 0.06$ arcsec.  Similar results are obtained when fitting a
Gaussian source on the image and then deconvolving from the
synthesized beam.
 
This source corresponds to galaxy ID=288391 of
\cite{scoville16}. However, they associated the ALMA flux with the
optical galaxy (see their Table B1) and measured the flux on an
aperture of up to 2.5'' radius centred on the galaxy position. This
explains the different flux reported by them.

\section{Optical and near-infrared photometry} \label{sec:phot}

To measure the redshift and physical properties of the ALMA source, we
extracted the photometry from the UltraVISTA-DR3 Y, J, H, K$_s$ bands
\citep{mccracken12}, from the HST I814 band \citep{koekemoer07}, and
from IRAC 3.6, 4.5, 5.0, 8.0\mic~ bands \citep{sanders07}.

As first step, we built the PSF of each image. For UltraVISTA and HST
we used a sample of bright, unsaturated stars. For IRAC, the PSFs were
obtained from synthetic instrument PSFs in the different channels
taking into account the contribution to the final mosaic from
observations at different position angles.

Secondly, for all bands we calibrated the RMS map and the background by
injecting fake sources in empty areas of the images as described in
\cite{merlin16}.

We performed source detection for the ALMA-detected galaxy on the K$_s$
band, where the faint source can be identified by visual inspection.
This band provides the best compromise between source brightness and
image resolution and allows us to separate the ALMA source from the
close-by optically bright extended galaxy.  Detection is obtained with
SExtractor \citep{bertin96} after optimizing the relevant parameters
in order to segment the area around our source in the most effective
way while keeping the two sources as separated.

Since the faint source lies on the tail of the brighter close-by
galaxy, the flux estimated by SExtractor may be
contaminated. Therefore, we estimated the total flux in all UltraVISTA
bands by adopting T-PHOT\footnote{ T-PHOT is a template-fitting
  photometry code developed within the ASTRODEEP project, designed to
  measure the photometry on low-resolution images by exploiting the
  prior information contained in images with higher resolution.}
\citep{merlin15} on an area of 60x60 arcsec, thus taking into account
the effect of source confusion.  We adopted the I814 cutouts as
high-resolution priors of the source light distribution, except for
the ALMA-detected galaxy that is undetected in the I814 image (see
below). For this source, we assumed a point-like object at the
position detected by SExtractor on the K$_s$-band image. From the
PSFs, we built convolution kernels between the I814 and the UltraVISTA
bands. These kernels were fed to T-PHOT to extract template-fitting
photometry.

We used a similar technique for the IRAC bands, where the two
sources are highly blended, but used K$_s$-band cutouts as priors.

To measure the total HST I814 flux, we first calculated a convolution
kernel by matching the I814 and the K$_s$-band PSFs. We then built a
version of the I814 image matched to the image in K$_s$ band, on which we
ran SExtractor.  We finally scaled the K$_s$-band total flux measured
with T-PHOT according to the colour measured with SExtractor in one
FWHM aperture (2 FWHM aperture for the more extended
optically detected galaxy) between these two images. 

We find that the ALMA source is well detected in all NIR bands, with a
S/N of $\sim$10, $\sim$16, $\sim$8 and $\sim$12 in the H, K$_s$, IRAC
CH1 and CH2 bands, respectively, $\sim$6 and $\sim$4 in Y and J, and
$\sim$2--3 in CH3 and CH4, while in the I814 band we obtain a
$1\sigma$ upper limit at AB$\sim$26.6. The measured fluxes for both
galaxies are given in Table \ref{tab:phot}.

We show in Fig.~\ref{fig:stamps} the postage stamps, in all available
images from I814 to radio wavelengths, where the ALMA and the
optical source are indicated. The ALMA-detected galaxy is not evident
in a by-eye inspection on the Y and J bands, possibly because of blending with
the brighter object, despite the relatively high S/N of the
detection. We therefore checked the covariance index estimated by
T-PHOT, that is, the ratio of the maximum covariance to the variance of
the object itself. As discussed in \cite{merlin15}, the covariance
index gives an idea of the reliability of the fit, with strongly
covariant objects (covariance index $\sim1$) that might be affected by
systematics. The covariance indexes for the Y and J bands are of the
order of $10^{-2}$, indicating that the fit is reliable and the
blending is not extreme and does not strongly affect the photometric
uncertainty.  Our detection is deeper than what is expected on the
basis of the limiting aperture magnitudes of the images reported by the
documentation attached to the UltraVISTA-DR3. This is possible since
T-PHOT estimates the photometry by weighting the source central region
more than the external (noisier) parts, at variance with an aperture
photometry giving equal weight to the entire extension of the source,
thus allowing for a better S/N.

\begin{table*}
\centering
\caption{Observed photometry of the ALMA-detected SMG and of the
  optically detected source.}  
\begin{tabular} {ccccc}
\hline 
\noalign{\smallskip} 
& & & lensed source & lens \\
& & & (ALMA-detected) & (optically detected) \\
\noalign{\smallskip} 
 Instrument & Filter & Central $\lambda$ & Flux &Flux \\
& & [$\mu$m] & [$\mu$Jy]& [$\mu$Jy] \\
\noalign{\smallskip} 
\hline 
\noalign{\smallskip} 
ACS/HST & I814W & 0.79 & $<$0.085 & 3.44 $\pm$ 0.12\\
VIRCAM/VISTA & Y & 1.02& 0.171 $\pm$ 0.028 & 5.726 $\pm$ 0.037\\
VIRCAM/VISTA & J & 1.25& 0.124 $\pm$ 0.030 & 7.871 $\pm$ 0.040\\
VIRCAM/VISTA & H & 1.65&0.455 $\pm$ 0.043 & 11.357 $\pm$ 0.059\\
VIRCAM/VISTA & K$_s$ & 2.15& 0.902 $\pm$ 0.057  & 17.600 $\pm$ 0.077\\
IRAC/{\it Spitzer} & CH1 & 3.56& 2.12 $\pm$ 0.27  & 24.2 $\pm$ 1.2\\
IRAC/{\it Spitzer} & CH2 & 4.51 &  4.33  $\pm$ 0.35  & 22.1 $\pm$ 1.5\\
IRAC/{\it Spitzer} & CH3 &5.74 & 6.8  $\pm$ 2.1  & 15.3 $\pm$ 8.5\\
IRAC/{\it Spitzer} & CH4 & 7.90 & 4.1  $\pm$ 2.4  & 30 $\pm$ 11\\
PACS/{\it Herschel}& 100$\mu$m & 101.74& \multicolumn{2}{c}{$<1.07 \cdot 10^3$} \\
PACS/{\it Herschel}$^a$& 160$\mu$m & 164.19 & $(5.46 \pm 3.79^b) \cdot 10^3$& $(5.34 \pm 3.79^b) \cdot 10^3$\\
SPIRE/{\it Herschel}$^a$ & 250$\mu$m & 251.89& $(13.92 \pm 3.45^b) \cdot 10^3$& $(8.11 \pm 3.45^b) \cdot 10^3$\\
SPIRE/{\it Herschel}$^a$ & 350$\mu$m & 351.92& $(19.78 \pm 4.10^b) \cdot 10^3$& $(8.72 \pm 4.10^b) \cdot 10^3$\\
SPIRE/{\it Herschel}$^a$ & 500$\mu$m & 509.81& $(10.03 \pm 4.76^b)  \cdot 10^3$& $(6.83 \pm 4.76^b)  \cdot 10^3$\\
ALMA & Band 7 & 873.37 &$(3.90 \pm 0.41) \cdot 10^3$ & --\\
VLA & Band L & $2 \cdot 10^5$ & 64 $\pm$ 11 & --\\
\noalign{\smallskip} \hline \noalign{\smallskip}
\end{tabular}
\tablefoot{
  \\
  The photometry of the ALMA-detected source has not been corrected for magnification.\\
  $^a$ Photometry in the 160, 250, 350 and 500 \mic~{\it Herschel} bands has
  been estimated through assumptions (see text) rather than directly measured.\\
  $^b$ The RMS value 
  includes confusion noise.
}
\label{tab:phot}
\end{table*}

\section{Physical properties} \label{sec:results}

We fit the photometry from the I814 band to IRAC with the PEGASE 2.0
templates \citep{fioc97} using our own code {\it zphot.exe}
\citep{fontana00,santini15}.  We infer a photometric redshift of 3.28
(Fig.~\ref{fig:photoz}).  The adoption of the \cite{bc03} library (see
below) instead of PEGASE 2.0 gives a very similar solution ($z$=3.25).
Although the $\chi^2$ curve is somewhat broad, a value for the
photometric redshift between 3 and 3.5 is also confirmed by the
starburst FIR-radio SED of \cite{yun02}, based on the location of the
FIR peak (see below for a derivation of the FIR fluxes), while the
lower and higher redshift solutions are excluded by the same template.

We note that the Y band scatters from the best fit by
$\sim$2$\sigma$. We do not expect strong emission lines at the
rest-frame wavelength corresponding to the observed Y band, unless the
true redshift is inconsistent with the inferred photo-z $\chi^2$
curve. Such a discrepancy between the observed flux and the best-fit
template, as well as the broadness of the $\chi^2$ curve, reflect the
difficulty in measuring the photometry in these complex situations (a
faint source on the tail of a bright one).

\begin{figure}[!t]
\resizebox{\hsize}{!}{\includegraphics[angle=270]{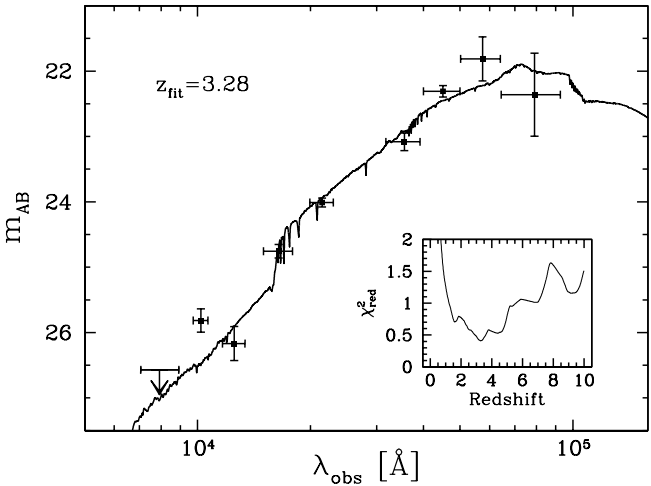}}
\caption{Best-fit with PEGASE 2.0 templates to infer the photometric
  redshift. The inset panel shows the reduced $\chi^2$.  }
\label{fig:photoz}
\end{figure}

\begin{figure}[!t]
\resizebox{\hsize}{!}{\includegraphics[angle=270]{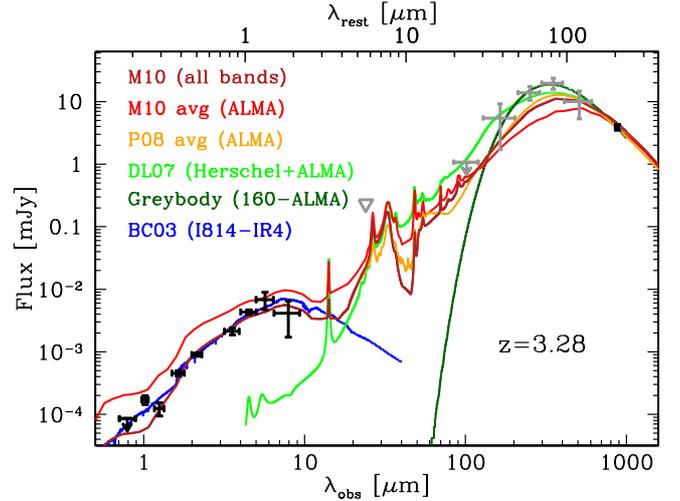}}
\caption{Best-fits of the ALMA-detected SMG. The black symbols show
  the measured fluxes, the grey symbols those estimated through
  assumptions in the {\it Herschel} bands (see text). The upside-down grey
  triangle shows the total 24\mic~ flux of the blended system.  The
  blue curve shows the best fit of the optical-to-NIR photometry with
  the \citet[BC03]{bc03} library; the brownish curve is the best fit
  of all available bands with the SMG templates of
  \cite{michalowski10}; the red and orange curves are the fit of the
  ALMA flux to the average SMG templates of \cite{michalowski10} and
  \cite{pope08}, respectively; the light green curve represents the
  best fit of the \cite{draineli07} library to the {\it Herschel} and ALMA
  bands; finally, the dark green curve is the fit with a modified
  blackbody with emissivity index $\beta=2$ to the 160\mic-to-ALMA
  bands.  See text for further details.}
 \label{fig:fits}
\end{figure}
 
With the inferred best-fit redshift, we fit the same photometric bands
with the \cite{bc03} library built by assuming exponentially
decreasing SF histories and adopting the very same assumptions in
terms of parameter grid, priors, etc. as detailed in Table 1 of
\citet[see column named Method 6a$_\tau$]{santini15}, except for the
metallicity that cannot be super-solar. We obtain a stellar mass of
$\sim 4\cdot 10^{10} M_\odot$.  The best fit is shown as a blue curve
in Fig.~\ref{fig:fits}, and the best-fit parameters are given in
Table~\ref{tab:physparam}.

Given the proximity of the two galaxies on the line of sight, it is
impossible to properly deblend the MIPS and {\it Herschel} fluxes (the FWHMs
range from $\sim$5 to $\sim$36 arcsec from 24\mic~to 500\mic). To
fully characterize the FIR-submm SED of the ALMA-detected SMG, we
started from the tentative assumption that the 24$\mu$m emission
($S_{24\mu m}=233\pm16$~$\mu$Jy) is completely associated with the
optically detected source, as the MIPS image is not deep enough to
detect galaxies at $z\sim3$. We assumed that the optical galaxy can
be described by the Main Sequence template of \cite{elbaz11}, as
confirmed by its best-fit stellar mass of
$\sim 2 \cdot 10^{10}M_\odot$ and SFR of $\sim 50 M_\odot/yr$ (the latter has been
  obtained by fixing this template to the
  observed 24\mic~ flux and adding the unobscured component following
  \citealt{santini09}). We estimated the
{\it Herschel}  fluxes of the optically detected galaxy by normalizing
the Main Sequence template of \cite{elbaz11} to the 24$\mu$m flux and
integrating on {\it Herschel}  filters. By subtracting the obtained values
from the {\it Herschel}  photometry of the blended galaxy system and adding
in quadrature an uncertainty of 35\% on the model
\citep[see][]{elbaz11}, we estimated the {\it Herschel}  fluxes of the ALMA
source. We then subtracted the 24$\mu$m flux predicted by fitting the
{\it Herschel}+ALMA bands of the ALMA-detected galaxy with the
\cite{draineli07} templates from the total value and repeated the
procedure, which converged after four iterations. The {\it Herschel}  fluxes
estimated for the ALMA-detected galaxy account for 51-69\% of the
total flux, depending on the band, while only less than 10\% of the
24$\mu$m is attributed to it. Since the system is undetected at
100\mic~in the public {\it Herschel}  catalogue, and this catalogue is cut at
3$\sigma$, we recomputed the photometry in this band with T-PHOT to
infer a 1$\sigma$ upper limit.

We can therefore build the full I814-to-submm SED of the ALMA source
and fit it with the SMG library of \cite{michalowski10}.  The best-fit
is shown as a brownish curve in Fig.~\ref{fig:fits}.  The inferred
total (8-1000 $\mu$m) IR luminosity, tracing the obscured SFR, is
$L_{IR}=6.02 \cdot 10^{12}L_\odot$.  Since  {\it Herschel} fluxes are
estimated through an indirect procedure and through assumptions that
are not necessarily verified, we verified that the result is unchanged
if we exclude  {\it Herschel} bands from the fit.  Furthermore, we fit the
ALMA flux with the average SMG templates of \cite{michalowski10} and
of \cite{pope08} (red and orange curves).  These provide $L_{IR}$ of
$3.27 \cdot 10^{12}L_\odot$ and $4.44 \cdot 10^{12}L_\odot$,
respectively. The dispersion of the above values is likely indicative
of the uncertainties associated with the total IR luminosity.  For
this reason, we decided to adopt the mean value and standard
dispersion as an estimate of the true $L_{IR}$ and its error bar. We
obtain $L_{IR} = (4.57 \pm 1.38) \cdot 10^{12} L_\odot$.

The total SFR is computed by adding the obscured and unobscured
components following the prescriptions adopted by \citet[see references therein]{santini09}, calibrated to the adopted IMF:
\begin{equation}
{\rm SFR_{IR+UV}}=10^{-10}\times L_{bol}/L_\odot
\end{equation}
\begin{equation}
 L_{bol} = 2.2 \times L_{UV} + L_{IR}
\end{equation}
where $L_{UV}=1.5 \times L_{2700\AA}$ is the rest-frame UV luminosity,
uncorrected for extinction, derived from the optical-to-NIR SED
fitting.  We infer $\rm SFR_{IR+UV} = 463 \pm 137 M_\odot/yr$.  As
expected for this type of sources, the obscured component
(${\rm SFR_{obsc}}/M_\odot yr^{-1}=10^{-10} \cdot L_{IR}/L_\odot$)
strongly dominates the total SFR ($\sim$99\%). Moreover, the total SFR
obtained from the UV and IR emission is 
higher than what is obtained by correcting the optical-to-NIR emission for
dust extinction (${\rm SFR_{SEDfit}}=121 M_\odot/yr$, although with
a huge uncertainty, inferred from the same fit to derive the stellar
mass), confirming the necessity of measuring the dust-enshrouded SFR to
achieve an unbiased view of the cosmic SFH, especially at these
redshifts (see e.g. \citealt{santini09}, \citealt{dunlop16}).

In the 20 cm VLA catalogue of \cite{bondi08} we found a radio
counterpart of our faint SMG at RA=10:01:38.558, DEC=+02:37:36.85,
consistent with the ALMA position. By adopting the prescriptions of
\cite{barger12}, the radio flux of $64 \pm 11$ $\mu$Jy translates into
an SFR estimate of $237 \pm 41 M_\odot/yr$.  According to
\cite{bell03}, the scatter in the FIR--radio correlation, upon which
the derivation is based, implies an uncertainty  of a factor of 2 in the
resulting SFR.

To search for an additional probe of the SFR, we looked for X-ray
counterparts in the {\it Chandra} catalogue of \cite{civano16} but found none.
Given the limiting depth of this catalogue, a non-detection implies a
0.5-10 keV flux lower than $8.9\times 10^{-16}$
erg~cm$^{-2}$~s$^{-1}$.

Finally, the fit with \cite{draineli07} templates (light green curve
in the figure) gives a dust mass of
$M_{dust}=1.12^{+0.16}_{-0.08} \cdot 10^9 M_\odot$, while the fit with
a modified blackbody (dark green curve) with emissivity index
$\beta=2$ and absorption cross section per unit dust mass at
240\mic~of 5.17 cm$^2$/g \citep{lidraine01,drainelee84} gives
$M_{dust}=5.97^{+0.13}_{-0.12} \cdot 10^8 M_\odot$ and
$T_{dust}=37^{+4}_{-3}K$ (the 100\mic~band is not used in the fitting
to avoid contamination from a hotter component,
\citealt{magnelli10}). The dust mass agrees with what is inferred from
the model of \cite{draineli07} after considering the scaling factor of
$\sim$1.5. Indeed, the attempt of reproducing the Wien side and at the
  same time the Rayleigh-Jeans side of the modified blackbody spectrum
  (instead of fitting a multi-temperature grain distribution) has the
  effect of overestimating the dust temperature and hence
  underestimating the dust mass, which is inversely proportional to the
  blackbody intensity (\citealt{santini14} and references therein;
\citealt{berta16}).  The best-fit temperature is in very good
agreement with the values $38^{+15}_{-5}$ and $43 \pm 10$ K found by
\cite{dacunha15} on the optically bright (they defined the
  subsample of SMGs detected in at least four optical/NIR bands as
  ``optically bright''; considering the different depth of the
  photometry used in the two works, our source would be classified as
  optically bright according to their criterion) and
$z$$>$2.7
subsamples, respectively.  As done for the SFR, we also computed the
dust mass from the ALMA flux only by assuming a temperature of
$\sim$40K
and obtained $M_{dust}=5.04
\pm 0.54 \cdot 10^8
M_\odot$, confirming the robustness of the result independently of the
{\it Herschel} photometry.  The inferred values for the dust mass support
the dust richness of SMGs with respect to their stellar content
compared to local star-forming galaxies and ULIRGs
\citep[e.g.][]{santini10,lofaro13,rowlands14,zavala15}.

As a sanity check, we also fit the full SED of our source with
CIGALE\footnote{http://cigale.lam.fr/} \citep{noll09}, which provides a
self-consistent estimate of the stellar mass, SFR and dust mass by
accounting for the energy balance between dust absorption and
re-emission. The Bayesian analysis gives $M_{star}=(4.92
\pm 0.80) \cdot 10^{10} M_\odot$, SFR $=(460 \pm 48)
M_\odot/yr$ and $M_{dust}=(1.32 \pm 0.29) \cdot 10^9
M_\odot$ ($M_{star}=(4.33 \pm 0.85) \cdot 10^{10} M_\odot$, SFR $=(366
\pm 88) M_\odot/yr$ and $M_{dust}=(2.21 \pm 1.57) \cdot 10^9
M_\odot$ excluding {\it Herschel} bands), consistent with the values
obtained above.

The physical properties inferred for the SMG studied in this work
agree with those reported by \cite{dacunha15} for their sample of SMGs
from the LESS survey \citep{weiss09} observed by ALMA.

\begin{table}
\centering
\caption{Observed physical properties of the lensed (ALMA-detected)
  and lensing (optically detected)
  galaxies. } 
\begin{tabular} {lcc}
\hline 
\noalign{\smallskip} 
& lensed source & lens \\
& (ALMA-detected) & (optically detected) \\
\noalign{\smallskip} 
\hline 
\noalign{\smallskip} 
 RA (J2000)&150.4106125&150.410339 \\
\noalign{\smallskip} 
 DEC (J2000)&$+$2.6268611 &$+$2.626380  \\
\noalign{\smallskip} 
$z$ & 3.28  {\it (phot.)}& 1.1458  {\it (spec.)}\\
\noalign{\smallskip} 
\ms~[\msun] &  $3.51^{+2.60}_{-1.39} \cdot 10^{10}$ & $1.85^{+0.19}_{-0.12}
                                              \cdot 10^{10}$  \\
\noalign{\smallskip} 
$\rm SFR_{SEDfit}$ [\msun/$yr$]& $121^{+141}_{-85}$ & $69^{+7}_{-4}$
  \\
\noalign{\smallskip} 
 $L_{IR} [L_\odot]$&$(4.57 \pm 1.38) \cdot 10^{12}$  &
                                                            $(4.28 \pm 0.32)\cdot 10^{11}$ \\
\noalign{\smallskip} 
 $\rm SFR_{IR+UV}$ [\msun/$yr$]&$463 \pm 137$  & $48.0 \pm 3.5$ \\
\noalign{\smallskip} 
 $\rm SFR_{radio}$ [\msun/$yr$] &$237\pm41$ & --\\
\noalign{\smallskip} 
\md~[\msun] &  $1.12^{+0.16}_{-0.08} \cdot 10^{9}$ & --\\
\noalign{\smallskip} 
$T_{dust}~[K]$ &  $37^{+4}_{-3}$ & --\\
\noalign{\smallskip} 
age  [Gyr] & $0.40^{+1.38}_{-0.27}$ & $0.32^{+0.08}_{-0.03}$ \\
\noalign{\smallskip} 
$\tau$ [Gyr] & $15.0^{+0.0}_{-14.9}$ & $1.0^{+14.0}_{-0.0}$\\
\noalign{\smallskip} 
E(B-V) [mag]& $0.55^{+0.10}_{-0.15}$ & $0.40^{+0.00}_{-0.00}$\\
\noalign{\smallskip} 
$Z~[Z_\odot]$& 1.0& 1.0\\
\noalign{\smallskip} 
$\mu$ &$1.54^{+0.13}_{-0.08}$& --\\
 \noalign{\smallskip} \hline \noalign{\smallskip}
\end{tabular}
\tablefoot{The physical properties of the ALMA-detected galaxy have not been corrected for magnification.
}
\label{tab:physparam}
\end{table}

\section{Discussion and conclusions} \label{sec:disc}

The $z$=3.28 SMG studied in the present work lies very close (2
arcsec) to the line of sight of a massive $z$$\sim$1
galaxy (see Sect.~\ref{sec:targetsel}).  The flux of the background
source is therefore likely to be boosted by gravitational lensing
effects, a phenomenon quite common in the submillimeter
\citep{negrello10,vieira13,weiss13}.

\begin{figure}[!t]
\resizebox{\hsize}{!}{\includegraphics[angle=90]{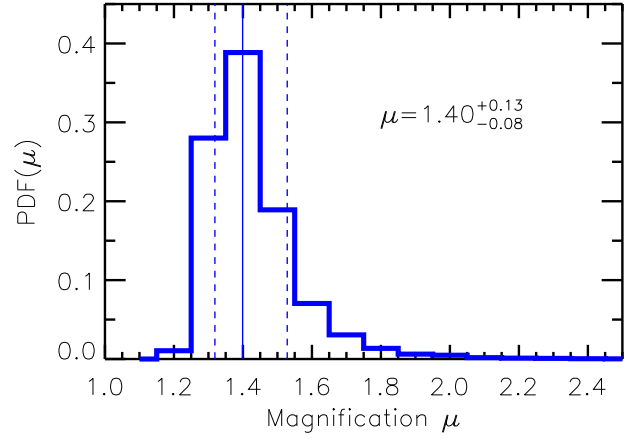}}
\caption{Distribution of the magnification factor arising from the
  uncertainty on the \cite{tully77} relation used to estimate the
  rotation velocity and hence the velocity dispersion of the lens,
  modelling it as a singular isothermal sphere (thick solid blue
  histogram, see text for details).  The solid and dashed black lines
  indicate the median and 68\% percentile range.  }
\label{fig:lens}
\end{figure}

The magnification factor arising from the lens has been estimated by
modelling the lens as a singular isothermal sphere \citep[][and
references therein]{mason15} where the velocity dispersion is
estimated via the rest-frame K-band \cite{tully77} relation inferred
by \cite{tiley16} at $z$$\sim$1
and dividing the rotation velocity by $\sqrt{2}$
\citep{kochanek04}. The estimated velocity dispersion for the lens is
$\sim$209$\pm$20 km/s.  The velocity dispersion scales with the dark
matter mass in the system and so is the best indicator of the strength
of the gravitational lens \citep{turner84,schneider06,treu10}.  The
uncertainty in the Tully-Fisher relation gives rise to a distribution
of magnification factors. The resulting distribution, shown in
Fig.~\ref{fig:lens}, has a median value of
$\mu=1.40^{+0.13}_{-0.08}$.  However, a singular isothermal sphere may
not be the best model for this lens, which seems to be elongated along
the SW-to-NE direction. According to \cite{barone-nugent15}, a
singular isothermal ellipsoid may be a more realistic
parameterization. Despite more complicate calculations, they claimed
that for low ellipticities ($\epsilon\lesssim 0.2$), as is the case of
our source (SExtractor estimates $\epsilon=0.18$ on the I814 band),
the magnification increases by $\simeq 10\%$ when the distant source
is located in the direction of the lens' major axis. We therefore
adopted a value of $1.54$ as an estimate of the magnification factor.

A magnification factor of 1.54 implies that the intrinsic SFR of our
SMG is between 154 and 300$M_\odot/yr$ (depending on the
chosen SFR tracer), a value consistent with that of the population of
faint SMGs that are important contributors to the cosmic SFH
\citep{ono14}.  Most importantly, being undetected in optical, this
galaxy is representative of the elusive dust-enshrouded star-forming
population that is typically not included in the measured SFH (nor in
the stellar mass density), even in a relatively well-studied and
deeply observed field such as COSMOS.  With an intrinsic K$_s$-band
magnitude of $\sim$24.5 and submm flux $S_{870\mu
  m}$$\sim$2.5~mJy,
this source is a factor of 2 fainter than the knee of the differential
number counts at 1.2 mm \citep{fujimoto16} (based on extrapolation
from the fits).  In particular, this is one of the still few faint
optically undetected SMGs (with an identified NIR counterpart) known
to date with full
SED characterization, which allows us not only to accurately estimate the
redshift but also to measure its stellar mass and other physical
properties (see e.g. \citealt{dacunha15} for a similar SED
characterization of the ALESS SMGs).  It adds one member to the 
still sparse population of known faint, dusty, star-forming galaxies.

Do the SF properties of the faint SMG studied in this work resemble
those of ``normal'' star-forming galaxies or those of extreme
starbursts?  To answer this question, we first consider the physical
size of our source.  At $z=3.28$,
the size measured on ALMA data translates into a physical size of
$1.8\times
1.6$ kpc.  This value is perfectly consistent with the average size
inferred by \cite{hodge16} for a sample of 16 bright
$z$$\sim$2.5
ALESS SMGs, whose formation, according to the authors, may be due to
major mergers. The measured size is larger than the $\sim$1 kpc size
inferred by \cite{simpson15} and \cite{ikarashi15} for bright high-$z$
SMGs, and smaller than 2.7 kpc, which is what is expected by
extrapolating the results of \cite{vanderwel14} for late-type galaxies
at the redshift and stellar mass of our source.  The compactness
compared to normal star-forming galaxies is another indication that
the properties of our faint SMGs resemble those of starbursts.  Where
is this source located with respect to the Main Sequence \citep[][and
references therein]{speagle14} of star-forming galaxies?  By assuming
the Main Sequence parameterization of \cite{schreiber15} (after converting
  stellar masses and SFRs from a Chabrier IMF into a Salpeter
 IMF 
by adding 0.24 dex, \citealt{santini12a},
  and 0.15 dex, \citealt{dave08}, respectively)  and that the
Main Sequence/starburst separation criterion by \cite{rodighiero11}
extends to $z$$\sim$3,
this galaxy would not be classified as a rare starburst, although it
lies on the upper envelope of the Main Sequence (close to the region
of starburst galaxies when the FIR-based SFR is adopted).  However,
its "starburstiness" $
R_{\rm SB}$, defined as the ratio between the specific SFR (${\rm SSFR
  =
  SFR}/M_{star}$) and the SSFR of a galaxy of the same mass located on
the Main Sequence ($R_{\rm
  SB}={\rm
  SSFR/SSFR_{MS}}$, \citealt{elbaz11}) is between 2 and 3.8 (depending
on the adopted SFR tracer). According to \cite{elbaz11},
starburstiness higher than 2 is indicative of a starburst nature.  A
different criterion for distinguishing starbursts from normal Main
Sequence galaxies is suggested by \cite{tan14} at high redshift
($z>4$),
based on the dust-to-stellar mass ratio: while this ratio is expected
to rise out to $z$$\sim$2.5
and then decrease for Main Sequence galaxies, starbursts are observed
to be more dust-rich at high redshift, providing evidence of an early
metal enrichment. The observed $M_{dust}/M_{star}$ of our source is
$\sim$0.03, in agreement with what has been observed by \cite{tan14} for
starburst galaxies at similar redshift and above the average ratio for
Main Sequence galaxies reported by \cite{bethermin15}.  All this is
consistent with the results of \cite{yamaguchi16}, who found that
while four out of the five faint SMGs of their sample are located in
the Main Sequence, the only source that is faint at optical and NIR
wavelengths is a starburst galaxy.

Observing even more galaxies like the one studied in this work is
of major importance to reach a full understanding of the population of
faint SMGs in the early Universe and for better constraining the
cosmic star formation history through a complete census of
star-forming galaxies. Indeed, at these redshifts, the latter is so
far almost completely and critically dependent on optical detections
and dust corrections.  The rapidly increasing number of observations
carried out with ALMA, as well as the advent of JWST for their
characterization, will be of great help in the near future.

\begin{acknowledgements}
  We thank the anonymous referee for the thorough review and
  helpful comments. PS thanks D. Burgarella and M. Boquien for
  support in using CIGALE and J. Dunlop for a critical review of the
  manuscript.  The research leading to these results has received
  funding from the European Union Seventh Framework Programme
  ASTRODEEP (FP7/2007-2013) under grant agreement n$^\circ$ 312725. RM and RA
  acknowledge support from the ERC Advanced Grant 695671 QUENCH. This
  paper makes use of the following ALMA data:
  ADS/JAO.ALMA\#2013.1.00034.S. ALMA is a partnership of ESO
  (representing its member states), NSF (USA) and NINS (Japan),
  together with NRC (Canada), NSC and ASIAA (Taiwan) and KASI
  (Republic of Korea), in cooperation with the Republic of Chile. The
  Joint ALMA Observatory is operated by ESO, AUI/NRAO and NAOJ. Based
  on data products from observations made with ESO Telescopes at the
  La Silla Paranal Observatory under ESO programme ID 179.A-2005 and
  on data products produced by TERAPIX and the Cambridge Astronomy
  Survey Unit on behalf of the UltraVISTA consortium.
\end{acknowledgements}

\bibliographystyle{aa}
\bibliography{biblio}

\end{document}